\documentstyle[12pt]{article}
\title{Signatures of discrete symmetries in the scalar sector}
\author{L.\ Lavoura\thanks{On leave of absence
from Universidade T\'ecnica de Lisboa,
Lisbon, Portugal} \\
\small Department of Physics, Carnegie-Mellon University, \\
\small Pittsburgh, Pennsylvania 15213, U.S.A.}
\begin{document}
\maketitle
\begin{abstract}
I discuss methods to identify the presence of dicrete symmetries
in the two-Higgs-doublet model
by observing the masses
and the cubic and quartic interactions of the scalars.
The symmetries considered are a $ Z_2 $ symmetry under which
$ \phi_2 \rightarrow - \phi_2 $,
and a CP symmetry which enforces real coupling constants
in the Higgs potential.
Those symmetries are spontaneously broken,
and the $ Z_2 $ symmetry may also be softly broken.
I identify the signatures in the interactions of the scalars
that these symmetries leave after their breaking.
\end{abstract}

\vspace{5mm}

Twenty-one years ago,
T.\ D.\ Lee \cite{lee}
pointed out that CP may be spontaneously broken.
In his two-Higgs-doublet model,
CP is a symmetry of the Lagrangian,
which is broken by the relative phase
between the vacuum expectation values (VEVs)
of the two Higgs doublets.
Other models of spontaneous CP violation
have been suggested since then \cite{branco,gerard},
and spontaneous CP violation has been used as an ingredient
in the building of many models \cite{nelson,babu}.
However,
no one has yet attempted to answer the following basic questions:
how can we experimentally distinguish between spontaneous
and explicit CP violation?
If CP violation is spontaneous,
does that fact lead to some relationships among the coefficients
of the various interaction terms in the Lagrangian,
relationships which might be experimentally tested for?
(At least in principle,
even if the practical measurements might be too difficult.)

In the context of the two-Higgs-doublet model,
it is usual to assume the existence of a discrete symmetry $ Z_2 $,
under which one of the two doublets changes sign,
while the other doublet remains unaffected.
That symmetry is softly broken in some models.
How can we assert experimentally whether such a symmetry exists or not,
and whether it is softly broken or not?

After discrete symmetries in the scalar sector
are spontaneously or softly broken,
do they still leave traces of their presence in the fundamental Lagrangian?

I present in this Brief Report a partial answer to these questions.

For definiteness,
I concentrate on the two-Higgs-doublet model.
That model is now very popular,
partly because two doublets is the Higgs structure
of the minimal supersymmetric standard model.
If there are more than two doublets
the algebra involved becomes extremely heavy.
I consider a SU(2)$ \otimes $U(1) gauge model with two scalar doublets
$ \phi_1 $ and $ \phi_2 $.
The most general Higgs potential consistent with renormalizability is
\begin{eqnarray}
V & = &
m_1 \phi_1^{\dagger} \phi_1 + m_2 \phi_2^{\dagger} \phi_2
+ (m_3 \phi_1^{\dagger} \phi_2 + h.c.)
\nonumber\\
  &   &
+ a_1 (\phi_1^{\dagger} \phi_1)^2
+ a_2 (\phi_2^{\dagger} \phi_2)^2
+ a_3 (\phi_1^{\dagger} \phi_1) (\phi_2^{\dagger} \phi_2)
+ a_4 (\phi_1^{\dagger} \phi_2) (\phi_2^{\dagger} \phi_1)
\nonumber\\
  &   &
+ \left[ a_5 (\phi_1^{\dagger} \phi_2)^2
+ a_6 (\phi_1^{\dagger} \phi_1) (\phi_1^{\dagger} \phi_2)
+ a_7 (\phi_2^{\dagger} \phi_2) (\phi_1^{\dagger} \phi_2)
+ h.c.\right]\, .
\label{eq:potentialphi}
\end{eqnarray}
All the coupling constants,
except $ m_3 $,
$ a_5 $,
$ a_6 $,
and $ a_7 $,
are real because of hermiticity.
I assume that the VEVs
of $ \phi_1 $ and $ \phi_2 $ are aligned,
in the sense that they preserve
the U(1) of electromagnetism.\footnote{T.\ D.\ Lee \cite{lee}
has shown that this happens if one inequality is satisfied
by the coupling constants of the potential.}
Those VEVs have a relative phase:
the VEV of $ \phi_1^0 $ is $ v_1 $,
and the VEV of $ \phi_2^0 $ is $ v_2 \exp (i \alpha) $,
$ v_1 $ and $ v_2 $ being real and positive.\footnote{The VEV of $ \phi_1^0 $
is made real and positive by a gauge transformation.
This represents no loss of generality.}
$ v = \sqrt{v_1^2 + v_2^2} $ is a measurable quantity,
$ v = 174 $ GeV.

Instead of working with $ \phi_1 $ and $ \phi_2 $,
it is convenient to work in the Georgi \cite{georgi} basis
of doublets $ H_1 $ and $ H_2 $,
with the following defining features:
$ H_1 $ has real and positive VEV $ v $,
while $ H_2 $ has vanishing VEV.
The Georgi basis is reached by means of the transformation
\begin{eqnarray}
\phi_1 & = & (v_1 H_1 + v_2 H_2) / v\, ,
\nonumber\\
\phi_2 & = & e^{i \alpha} (v_2 H_1 - v_1 H_2) / v\, .
\label{eq:changeofbasis}
\end{eqnarray}
In the Georgi basis,
the Higgs potential reads
\begin{eqnarray}
V & = &
\mu_1 H_1^{\dagger} H_1 + \mu_2 H_2^{\dagger} H_2
+ (\mu_3 H_1^{\dagger} H_2 + h.c.)
\nonumber\\
  &   &
+ \lambda_1 (H_1^{\dagger} H_1)^2
+ \lambda_2 (H_2^{\dagger} H_2)^2
+ \lambda_3 (H_1^{\dagger} H_1) (H_2^{\dagger} H_2)
+ \lambda_4 (H_1^{\dagger} H_2) (H_2^{\dagger} H_1)
\nonumber\\
  &   &
+ \left[ \lambda_5 (H_1^{\dagger} H_2)^2
+ \lambda_6 (H_1^{\dagger} H_1) (H_1^{\dagger} H_2)
+ \lambda_7 (H_2^{\dagger} H_2) (H_1^{\dagger} H_2)
+ h.c.\right]\, ,
\label{eq:potentialH}
\end{eqnarray}
in which all the coupling constants,
except $ \mu_3 $,
$ \lambda_5 $,
$ \lambda_6 $,
and $ \lambda_7 $,
are real by hermiticity.
Because only $ H_1 $
has a non-zero VEV,
$ v $,
which is real,
the stationarity conditions of the vacuum read
\begin{eqnarray}
\mu_1 = - 2 \lambda_1 v^2\, ,
\label{eq:mu1}\\
\mu_3 = - \lambda_6 v^2\, .
\label{eq:mu3}
\end{eqnarray}
I use these conditions to eliminate $ \mu_1 $ and $ \mu_3 $
as independent variables from $ V $.
Because $ \mu_3 $ is complex while $ \mu_1 $ is real,
Eqs.~\ref{eq:mu1} and \ref{eq:mu3}
constitute three real equations.
They correspond to the three real equations which,
in the basis of $ \phi_1 $ and $ \phi_2 $,
determine the stability of the vacuum by fixing the partial derivatives
of the vacuum potential with respect to $ v_1 $,
$ v_2 $,
and $ \alpha $,
to be zero.

The Georgi basis is useful because
the Goldstone modes are perfectly identified when one uses it.
Writing
\begin{eqnarray}
H_1 & = &
\left( \begin{array}{c}
G^+ \\ v + (H^0 + i G^0)/\sqrt{2}
\end{array} \right)\, ,
\label{eq:H1}\\
H_2 & = &
\left( \begin{array}{c}
H^+ \\ (R + i I)/\sqrt{2}
\end{array} \right)\, ,
\label{eq:H2}
\end{eqnarray}
$ G^+ $ and $ G^0 $ are the Goldstone bosons which,
in the unitary gauge,
become the longitudinal components of the $ W^+ $ and of the $ Z^0 $.
$ H^0 $,
$ R $ and $ I $ are real neutral fields,
which are linear combinations
of the three physical scalars $ X_k $,
\begin{equation}
\left( \begin{array}{c} X_1 \\ X_2 \\ X_3 \end{array} \right)
=
T\, \left( \begin{array}{c} H^0 \\ R \\ I \end{array} \right)\, ,
\label{eq:physicalscalars}
\end{equation}
$ T $ being an orthogonal matrix.
$ H^+ $ is the physical charged scalar.
As a consequence of this,
it is easy to write down the mass terms \cite{silva}
and the cubic and quartic interactions of the physical scalars
as functions of the coupling constants of the potential in the Georgi basis,
Eq.~\ref{eq:potentialH}.
By observation of those masses and interactions,
$ \mu_2 $ and the $ \lambda_i $ ($ i $ from 1 to 7)
can be measured.\footnote{Once $ \mu_2 $ and the $ \lambda_i $
have been measured,
one can get from them the coupling constants
of the potential in the original basis,
Eq.~\ref{eq:potentialphi},
and also $ v_1/v_2 $ and $ \alpha $.
But the quantities more directly measured
are $ \mu_2 $ and the $ \lambda_i $.}
Because $ \lambda_5 $,
$ \lambda_6 $ and $ \lambda_7 $ are complex,
this corresponds to a total of eleven real quantities.

However,
the Georgi basis is not totally well defined:
$ H_2 $ can suffer a U(1) rephasing,
while preserving its defining property of having a zero VEV.
When this is done,
the phases of $ \lambda_6 $ and of $ \lambda_7 $
get changed by the arbitrary phase $ \varphi $,
while the phase of $ \lambda_5 $ gets changed by $ 2 \varphi $.
Only rephasing-invariant combinations are measurable.
Therefore,
there are only ten real measurable quantities
in the scalar masses and interactions.
Two of these are phases connected to the presence of CP violation \cite{silva}.
For instance,
the phases of $ \lambda_6^{\ast} \lambda_7 $
and of $ \lambda_5^{\ast} \lambda_6 \lambda_7 $ are
two independent measurable phases,
while the phase of $ \lambda_5^{\ast} \lambda_7^3 $ is not measurable.
The other eight measurable quantities are the moduli of $ \mu_2 $
and of the $ \lambda_i $.

Let us consider exactly how
the various parameters of the potential in the Georgi basis
might be measured.
Let us denote by $ A_k $ ($ k $ from 1 to 3)
the squared masses of the three neutral scalars $ X_k $.
Those squared masses are the eigenvalues of the mass matrix $ M $
in the basis $ (H^0, R, I) $,
which mass matrix has been explicitly written down in \cite{silva}.
I denote by $ A_+ $ the squared mass of the charged scalar $ H^+ $.
The only observables in the matrix $ T $
are the matrix elements of its first column \cite{silva}.
For instance,
the $ Z^0_{\mu} $ couples to the following current:
\begin{equation}
\frac{g}{2 \cos \theta_W} \left[
\sum_{k=1}^3 T_{k1} \left(
X_k \partial^{\mu} G^0 - G^0 \partial^{\mu} X_k \right)
+ \sum_{k,l,m} \epsilon_{klm} T_{k1} X_l \partial^{\mu} X_m
\right]\, ,
\label{eq:Zcouplings}
\end{equation}
$ \epsilon_{klm} $ being the totally antisymmetric tensor
with $ \epsilon_{123} = 1 $.\footnote{I omitted in the current
in Eq.~\ref{eq:Zcouplings} the terms involving the charged scalars.}
One should remember that $ T_{11}^2 + T_{21}^2 + T_{31}^2 = 1 $
because of the orthogonality of $ T $,
therefore only two of the three $ T_{k1} $ are independent.
Once one knows $ A_1 $,
$ A_2 $,
$ A_3 $,
$ A_+ $,
and the three $ T_{k1} $,
one can find the values of six parameters of the potential
via the following equations:
\begin{equation}
A_+ = \mu_2 + \lambda_3 v^2\, ,
\label{eq:k6}
\end{equation}
\begin{equation}
M_{11} = 4 \lambda_1 v^2 = \sum_{k=1}^3 A_k T_{k1}^2\, ,
\label{eq:k1}
\end{equation}
\begin{equation}
M_{22} + M_{33} = 2 \left( \frac{\mu_2}{v^2}
+ \lambda_3 + \lambda_4 \right) v^2 = \sum_{k=1}^3 A_k - M_{11}\, ,
\label{eq:k2}
\end{equation}
\begin{equation}
M_{22} M_{33} - M_{23}^2 =
\left[ \left( \frac{\mu_2}{v^2}
+ \lambda_3 + \lambda_4 \right)^2 + 4 |\lambda_5|^2 \right] v^4
= A_1 A_2 T_{31}^2 + A_1 A_3 T_{21}^2 + A_2 A_3 T_{11}^2\, ,
\label{eq:k3}
\end{equation}
\begin{equation}
M_{12}^2 + M_{13}^2 = 4 |\lambda_6|^2 v^4
= \sum_{k=1}^3 A_k^2 T_{k1}^2 ( 1 - T_{k1}^2 )
- 2 \sum_{k<l} A_k A_l T_{k1}^2 T_{l1}^2\, ,
\label{eq:k4}
\end{equation}
\begin{eqnarray}
2 M_{12} M_{13} M_{23} - M_{22} M_{13}^2 - M_{33} M_{12}^2
+ M_{11} (M_{22} + M_{33})
& = &
\nonumber\\
4 v^6 \left[ 2 {\rm Re} \left( \lambda_5^{\ast} \lambda_6^2 \right)
- \left( \frac{\mu_2}{v^2} + \lambda_3 + \lambda_4 \right)
|\lambda_6|^2 \right]
+ M_{11} (M_{22} + M_{33})
& = & A_1 A_2 A_3\, .
\label{eq:k5}
\end{eqnarray}
The other four parameters of the potential might be determined
in the following way.
The following cubic interactions are present in the potential:
\begin{equation}
V = \sqrt{2} v H^- H^+ \left( \lambda_3 H^0 + R {\rm Re} \lambda_7
- I {\rm Im} \lambda_7 \right) + ...
\label{eq:cubic1}
\end{equation}
By diagonalizing the mass matrix $ M $ we find that this interaction
is written in terms of the eigenstates of mass $ X_1 $,
$ X_2 $ and $ X_3 $ as
\begin{eqnarray}
  &   & \sqrt{2} v H^- H^+ \left( \lambda_3 H^0 + R {\rm Re} \lambda_7
- I {\rm Im} \lambda_7 \right)
\nonumber\\
  & = & \sqrt{2} v H^- H^+ \sum_{k=1}^3 X_k \left[ \lambda_3 T_{k1}
+ \frac{4 v^4
{\rm Re} \left( \lambda_5^{\ast} \lambda_6 \lambda_7 \right)}{T_{k1}
(A_k - A_l) (A_k - A_m)} \right.
\nonumber\\
  &  & + \left. \frac{v^2}{T_{k1}}
{\rm Re} \left( \lambda_6 \lambda_7^{\ast} \right)
\left( \frac{1 - T_{l1}^2}{A_k - A_m}
+ \frac{1 - T_{m1}^2}{A_k - A_l} \right) \right]\, ,
\label{eq:cubic2}
\end{eqnarray}
with $ l \neq k $ and $ m \neq k $ and $ l \neq m $.
This allows us to find the values of $ \lambda_3 $,
$ {\rm Re} \left( \lambda_6 \lambda_7^{\ast} \right) $,
and $ {\rm Re} \left( \lambda_5^{\ast} \lambda_6 \lambda_7 \right) $.
Finally,
$ \lambda_2 $ can be found,
for instance,
from the fact that it is the coefficient
of the $ (H^- H^+)^2 $ quartic interaction.

Suppose that there is an exact symmetry $ Z_2 $ under which
$ \phi_1 \rightarrow \phi_1 $ while
$ \phi_2 \rightarrow - \phi_2 $.
Then,
$ m_3 = 0 $ and $ a_6 = a_7 = 0 $.
Because there is then only one term in the potential which sees the
relative phase of $ \phi_1 $ and $ \phi_2 $
[the term $ a_5 (\phi_1^{\dagger} \phi_2)^2 + h.c $],
there is no CP violation \cite{gerard,noCP}.
$ a_5 $ can be set real by a rephasing of $ \phi_2 $.
In the case of exact $ Z_2 $ symmetry
there are therefore seven parameters in the Higgs potential:
$ m_1 $,
$ m_2 $,
$ a_1 $,
$ a_2 $,
$ a_3 $,
$ a_4 $ and $ a_5 $.
These seven parameters determine $ v $ and the ten coefficients
in the scalar masses and cubic and quartic interactions.
We thus expect four predictions.
Two of those predictions are connected to the absence of CP violation
in this model:
the two independent measurable phases vanish.
The other two predictions can be derived by the following method.
One first uses the stability conditions of the vacuum
to write $ m_1 $ and $ m_2 $ as functions of $ v_1 $ and of $ v_2 $,
or,
equivalently,
of $ v $ and the ratio $ v_1 / v_2 $.
One then uses the change of basis in Eq.~\ref{eq:changeofbasis}
to write $ \mu_2 / v^2 $ and the seven $ \lambda_i $
(which in this case are all real)
as functions of $ v_1 / v_2 $ and of $ a_1 $,
$ a_2 $,
...,
$ a_5 $
(being dimensionless,
they cannot depend on $ v $).
One finally inverts those equations to find
$ v_1 / v_2 $ and $ a_1 $,
$ a_2 $,
...,
$ a_5 $ as functions of $ \mu_2 / v^2 $
and of the $ \lambda_i $,
in this process obtaining the following two relationships among the
measurable parameters:
\begin{eqnarray}
\frac{\mu_2}{v^2} (\lambda_6 + \lambda_7)
+ 2 (\lambda_1 \lambda_7 + \lambda_2 \lambda_6)
& = & 0\, ,
\label{eq:relacao11}\\
\left( \frac{\mu_2}{v^2} + \lambda_3 + \lambda_4 + 2 \lambda_5 \right)
(\lambda_2 - \lambda_1)
+ \lambda_6^2 - \lambda_7^2
& = & 0\, .
\label{eq:relacao12}
\end{eqnarray}
These equations are the answer to the questions addressed in this Brief Report
for the particular case of the model with non-softly-broken
$ Z_2 $ symmetry.
After measuring $ \mu_2 / v^2 $ and the seven $ \lambda_i $
by observing the scalar masses and the scalar cubic and quartic interactions,
one should check whether there is no CP violation,
and whether Eqs.~\ref{eq:relacao11} and \ref{eq:relacao12} are satisfied.
If this applies,
we are in the presence of a model with non-softly-broken $ Z_2 $
symmetry.

I now address the case
of softly broken $ Z_2 $ symmetry
and spontaneously broken CP symmetry \cite{georgi2,eu,moorhouse}.
In this case,
$ a_6 $ and $ a_7 $ are zero because of the $ Z_2 $ symmetry.
$ m_3 $ is non-zero,
breaking the $ Z_2 $ symmetry softly.
However,
there is CP invariance in the Higgs potential,
$ m_3 $ and $ a_5 $ are real.
CP is broken by the phase $ \alpha $
between the VEVs of $ \phi_1 $ and $ \phi_2 $.
We now have ten measurable quantities written as functions of seven parameters
($ a_1 $,
$ a_2 $,
...,
$ a_5 $,
$ \alpha $ and $ v_1 / v_2 $).
We expect three equations among the observable quantities to hold.
After some work we find them to be
\begin{equation}
{\rm Im} \left[ \lambda_5^{\ast} (\lambda_6 + \lambda_7)^2 \right]
+ (\lambda_2 - \lambda_1) {\rm Im} (\lambda_6 \lambda_7^{\ast}) = 0\, ,
\label{eq:relacao21}
\end{equation}
and
\begin{equation}
(\lambda_2 - \lambda_1) {\rm Im} \left[ \lambda_5^{\ast}
(\lambda_6^2 - \lambda_7^2) \right]
- \left[ \lambda_2^2 - \lambda_1^2 -
(\lambda_2 - \lambda_1) (\lambda_3 + \lambda_4)
+ |\lambda_7|^2 - |\lambda_6|^2 \right]
{\rm Im} (\lambda_6 \lambda_7^{\ast}) = 0\, ,
\label{eq:relacao22}
\end{equation}
and
\begin{eqnarray}
\left[
\left( \frac{\mu_2}{v^2} \right)^2 (\lambda_1 - \lambda_2)
+ 2\, \frac{\mu_2}{v^2}\,
(\lambda_3 + \lambda_4) (\lambda_1 - \lambda_2) \right.
&   &
\nonumber\\
+ \left. \left( \frac{\mu_2}{v^2} + \lambda_1 + \lambda_2 \right) f_2
+ (2 \lambda_3 + 2 \lambda_4 - \lambda_1 - \lambda_2 )
( \lambda_1^2 - \lambda_2^2)
\right] (f_1)^2
&   &
\nonumber\\
+ (\lambda_1 - \lambda_2) \left[ f_2^2 - 4 f_3^2
+ 2 (\lambda_1 - \lambda_2) (\lambda_3 + \lambda_4 - \lambda_1 - \lambda_2)
\right] f_1
&   &
\nonumber\\
- (\lambda_1 - \lambda_2)^3 \left[ (f_2)^2 + 4 (f_3)^2 \right]
& = & 0\, ,
\label{eq:relacao23}
\end{eqnarray}
where
\begin{eqnarray}
f_1 & \equiv &
|\lambda_6|^2 + |\lambda_7|^2 + 2 {\rm Re} (\lambda_6 \lambda_7^{\ast})\, ,
\label{eq:f1}\\
f_2 & \equiv &
|\lambda_7|^2 - |\lambda_6|^2\, ,
\label{eq:f2}\\
f_3 & \equiv & {\rm Im} (\lambda_6 \lambda_7^{\ast})\, .
\label{eq:f3}
\end{eqnarray}

Let us now consider the Lee model of spontaneous CP violation.
In that model,
there is no $ Z_2 $ symmetry,
and therefore $ a_6 $ and $ a_7 $ are non-zero,
but there is CP invariance at the Lagrangian level,
which means that $ m_3 $,
$ a_5 $,
$ a_6 $ and $ a_7 $ are real.
Therefore,
the difference from the previous model is the existence of two extra
real couplings,
$ a_6 $ and $ a_7 $.
However,
$ \phi_1 $ and $ \phi_2 $,
because there is now no symmetry $ Z_2 $ which distinguishes between them,
are not uniquely defined.
One may rotate $ \phi_1 $ and $ \phi_2 $ freely
by means of an orthogonal transformation
(orthogonal and not unitary,
because we want to preserve a real scalar potential),
\begin{equation}
\left( \begin{array}{c} \phi_1 \\ \phi_2 \end{array} \right)
\rightarrow
\left( \begin{array}{cc} \cos \theta & \sin \theta \\
- \sin \theta & \cos \theta \end{array} \right)
\left( \begin{array}{c} \phi_1 \\ \phi_2 \end{array} \right)\, ,
\label{eq:rotation}
\end{equation}
in such a way as to eliminate one of the parameters of the potential,
$ m_3 $ for instance \cite{lee}.
There is thus one parameter in the potential which is spurious.
Therefore,
the two extra parameters $ a_6 $ and $ a_7 $
really correspond to only one extra degree of freedom.
We thus expect two relations
(instead of three as in the previous model)
to hold among the measurable quantities.
Indeed,
Eqs.~\ref{eq:relacao21} and \ref{eq:relacao22}
do not hold any more.
However,
one linear combination of them still holds when $ a_6 $ and $ a_7 $
are not zero:
\begin{eqnarray}
\left[ (\lambda_2 - \lambda_1) \left( \lambda_3 + \lambda_4
+ \frac{\mu_2}{v^2} \right) + |\lambda_6|^2 - |\lambda_7|^2 \right]
{\rm Im} ( \lambda_6 \lambda_7^{\ast} )
&   &
\nonumber\\
+ \frac{\mu_2}{v^2} {\rm Im}
\left[ \lambda_5^{\ast} (\lambda_6 + \lambda_7)^2 \right]
+ 2 \lambda_2 {\rm Im} \left[ \lambda_5^{\ast} \lambda_6
(\lambda_6 + \lambda_7 ) \right]
+ 2 \lambda_1 {\rm Im} \left[ \lambda_5^{\ast} \lambda_7
(\lambda_6 + \lambda_7 ) \right]
& = & 0\, .
\label{eq:relacao31}
\end{eqnarray}
Eq.~\ref{eq:relacao23} does not hold in the Lee model either.
I expect one further constraint among the physical observables
to hold in the Lee model,
which should somehow involve the quantity in the left-hand side
of Eq.~\ref{eq:relacao23}.
Unfortunately,
I have been unable to find this extra condition
characteristic of the Lee model.

As a final model,
let us now assume that CP violation is explicit,
not spontaneous,
but that there is a softly broken $ Z_2 $ symmetry.
That is,
$ a_6 = a_7 = 0 $,
while $ m_3 $ and $ a_5 $ are complex.
However,
because we are now free to rephase $ \phi_2 $,
it is only the phase of $ m_3^{\ast} a_5^2 $ which is relevant.
Therefore,
this model has only one more degree of freedom
than the model with softly broken $ Z_2 $ symmetry
and spontaneous CP violation.
We therefore expect two conditions
among the physical observables.
Indeed,
one finds that in this case Eqs.~\ref{eq:relacao21} and \ref{eq:relacao22}
hold,
but Eq.~\ref{eq:relacao23} does not.

I summarize my results.
Any two-Higgs-doublet model can conveniently be written
in the Georgi basis.
The parameters of the potential in that basis constitute a set
of ten independent quantities,
which can be directly measured by considering the coefficients
of the various scalar cubic and quartic interactions,
and the scalar masses.
Those ten quantities are $ \mu_2 $,
$ \lambda_1 $,
$ \lambda_2 $,
$ \lambda_3 $,
$ \lambda_4 $,
$ |\lambda_5| $,
$ |\lambda_6| $,
$ |\lambda_7| $,
and two independent CP-violating phases,
the phases of $ \lambda_6 \lambda_7^{\ast} $
and of $ \lambda_5^{\ast} \lambda_6 \lambda_7 $,
for instance.
If some discrete symmetries,
like a $ Z_2 $ symmetry or CP symmetry,
are imposed on the potential in a non-Georgi basis,
then some equations will hold among these ten
otherwise independent quantities.
In a model with non-softly-broken (but spontaneously broken)
$ Z_2 $ symmetry,
there is CP conservation
(which means that the two physical phases vanish),
and the conditions of Eqs.~\ref{eq:relacao11}
and \ref{eq:relacao12} hold.
If $ Z_2 $ is softly broken,
but CP is a spontaneously-broken symmetry of the potential,
Eqs.~\ref{eq:relacao21},
\ref{eq:relacao22} and \ref{eq:relacao23} hold.
If $ Z_2 $ is softly broken,
and CP is explicitly broken,
Eqs.~\ref{eq:relacao21} and \ref{eq:relacao22} hold,
but Eq.~\ref{eq:relacao23} does not.
It is interesting to observe that the difference between these two cases
is only whether Eq.~\ref{eq:relacao23} holds or not;
as Eq.~\ref{eq:relacao23} is extremely complicated,
in practice it will certainly be impossible
to distinguish between the two models in this way.
Finally,
in the general Lee model of spontaneous CP violation,
without any $ Z_2 $ symmetry,
Eq.~\ref{eq:relacao31} holds and,
presumably,
another much more complicated condition will also hold.

It is fair to say that all the conditions found are rather complicated
and should be very difficult to test.
This unfortunate result means that,
possibly,
much theoretical speculation on the existence of discrete symmetries
in the scalar scetor may be untestable in practice.
Also note that those conditions are just tree-level ones and,
because the symmetries on which they depend are broken,
should receive finite radiative corrections.

\vspace{2mm}

I thank Jo\~ao P.\ Silva for useful discussions,
and for collaboration in the early stages of this research.
Lincoln Wolfenstein and Ling-Fong Li read and criticized the manuscript.
This work was supported by the United States Department of Energy,
under the contract DE-FG02-91ER-40682.

\vspace{5mm}

%
%
\end{document}